# Navigating Selection Bias in Genome-wide Association Studies (GWAS)

Ciarrah-Jane S Barry*[1,2], Tahereh Gholipourshahraki[1,3], Zoltán Kutalik[4,5,6], Bjarni J Vilhjálmsson[1,2,3], Esben Agerbo[1,2]

[1] National Centre for Register-based Research, Aarhus University, Aarhus, Denmark

[2] Pioneer Centre for SMARTbiomed, Aarhus University, Aarhus, Denmark

[3] Bioinformatics Research Centre (BIRC), Aarhus University, Aarhus, Denmark

[4] University Center for Primary Care and Public Health, Lausanne, Switzerland

[5] Department of Computational Biology, University of Lausanne, Lausanne, Switzerland

[6] Swiss Institute of Bioinformatics, Lausanne, Switzerland

*Correspondence: ciarrah.barry@ph.au.dk (C.S.Barry), +45 8716 7936

Address: National Centre for Register-based Research, Bartholins Alle 2, Building 1260, DK-8000 Aarhus C, Denmark

## Abstract

Selection bias arises when inclusion in a genetic study depends on traits, or causes of traits, that are associated with genotype. It is present in volunteer-based biobanks, case-control designs and studies of disease progression, and genome-wide association study (GWAS) summary statistics from affected samples can propagate this bias into post-GWAS analyses, such as polygenic scores (PGS), Mendelian randomization (MR) and genetic correlation analyses. Despite broad potential impact of selection bias it is sometimes ignored in genetic studies. Many methods exist to detect and correct selection bias, but they rely on different assumptions and require specific data. We organise ten methods applicable to GWAS by the decisions an analyst faces: whether to diagnose or correct, which data are available (individual-level or summary-level), and which form of selection is operating (non-representative participation, case-control ascertainment, index events, sex-differential participation or survival to recruitment). For each method, we describe its principle, applicability, limitations and data requirements. A companion GitHub repository implements nine methods in R and prepares input for the external SPREG software. It provides worked examples using simulated data with known true effects for each form of selection. An interactive tool supports method selection. We conclude with open methodological challenges.

## Key points

- Selection bias in GWAS can appear in distinct forms due to multiple reasons (non-representative participation, case-control ascertainment, index events, sex-differential participation and survival to recruitment), each requiring specific correction.

- The choice of correction depends both on what causes selection and on the data available, for example inverse probability weighting requires selection explained by measured factors.

- Summary-level corrections for index event bias differ mainly in their assumptions about pleiotropy, and comparing methods with different assumptions indicates whether conclusions depend on any one of them.

- Our companion R repository, worked examples on simulated data with known true effects, and an interactive SHINY tool support user implementation of the methods.

# Introduction

Linking large-scale genetic biobanks to electronic health records and phenotypic information has rapidly transformed genetic research, enabling genome-wide association studies (GWAS) of thousands of outcomes in millions of participants from diverse populations (1). These resources, such as the UK Biobank (UKB) and Million Veteran Program, have advanced our understanding of the genetic basis of health outcomes, complex traits and molecular measurements (2,3). They have also informed research on causal relationships in health, helped identify potential drug targets and supported diagnostic development (2–5).

However, biobank participation processes are usually non-random associated with individual characteristics and impacted by loss to follow-up and survival (6–9). Other potential sources of non-random sampling include family-based and geographic recruitment patterns that may correlate with ancestry but are not equivalent to classical population stratification bias (7,10,11). When participation, follow-up or survival depend on genetic variation or phenotypic characteristics, single nucleotide polymorphism (SNP)-trait associations can be distorted. This is selection bias: the study sample is not representative of the target population, and the sampled group is not exchangeable with it.

Empirical studies have demonstrated its pervasiveness (12–17). For example, Danish national register-based data show substantial psychiatric multimorbidity among individuals with schizophrenia, with distinct trajectories across risk-factor profiles (18). Yet, analyses in UKB, a volunteer-based cohort, demonstrate systematically attenuated clinical and genetic severity, including lower polygenic scores (PGS), milder symptom profiles and more favourable outcomes (19). Together, these findings show that participation-based sampling can alter observed phenotypic and genetic distributions among individuals with schizophrenia. These differences can affect inference about disease architecture and genetic effects. They may lead to misidentified disease risk factors, distorted treatment effect estimates and biased individual risk estimates, with consequences for clinical decisions, research priorities and public health recommendations (20,21).

Formally, as defined by Hernán and Robins, *selection bias* refers to the distortion of an estimated association between an exposure and outcome that occurs when an individual's inclusion in the sample is not independent of the exposure and outcome, or factors that cause them (22). This may compromise both internal validity, inducing biased associations within the selected sample, and external validity, limiting the generalisability or transportability of findings to the target population (23). In GWAS the exposure is genotype and the estimand is the SNP–trait association in a defined target population. Non-representative sampling and collider bias are essentially two versions of the same problem: participation that depends on the trait distorts sample distributions (external validity), and participation that depends on both a SNP-associated factor and the trait induces spurious associations within the sample (internal validity). This review covers both. We do not address the counterfactual definitions of selection bias used for causal treatment-effect estimands, which are discussed elsewhere (22). Mechanisms of selection bias affecting sample composition are outlined in Box 1.

*Box 1: Mechanisms of selection affecting sample composition, with examples and potential implications for downstream analyses (24–26).*

**Healthy volunteer effect/Non-random participation/Participation bias**

Non-random participation can produce unrepresentative samples of the target population. Certain groups, such as people who are homeless, severely ill or institutionalised, may be systematically excluded due to barriers to participation. In UKB, a volunteer-based sample, participants are older, slightly more often female, healthier, and of higher socioeconomic position than the general UK population (8). This limits the generalisability of sample estimates and predictive models derived from the data. However, within-sample associations may remain valid, provided the selection mechanism is independent of the SNP-outcome relationship, given measured covariates.

**Attrition/Non-response/Sampling bias**

Exposure- or outcome-related factors may influence participation, follow-up, or drop-out, leading to non-random missingness and a non-representative sample. For example, in the Millennium Cohort Family Study, age, sex and ethnicity were associated with survey non-response (27).

**Immortal time bias/Survivor bias**

Individuals must survive to a specific point (e.g., variable measurement, recruitment) to participate. Implicitly healthier individuals are selected, potentially overestimating protective effects, as seen in early pharmaco-epidemiological studies (28,29). This has been framed as an emulation failure, where selection bias due to survival until exposure allocation induces confounding by unmeasured factors determining both survival and/or the outcome (30). In GWAS, variants that shorten life are under-represented among older participants, which can bias associations with late-onset traits.

**Competing risk**

A competing event (e.g., death) precludes the outcome of interest and may prevent inclusion in the sample (31,32). If inclusion requires remaining free of a competing event until recruitment or follow-up, conditioning on this requirement may induce selection bias.

**Ascertainment bias/Collider bias/Index-event bias (IEB)**

A variable causally influenced by both exposure and outcome is a collider. The probability of sample participation may be associated with a collider. Conditioning upon a collider, through adjustment or selection, induces bias in associations. For example, when a sample is enriched or depleted for a disease, SNPs that raise disease risk become associated with other causes of that disease. This can inflate, attenuate or reverse effect estimates. In UKB, associations between type 2 diabetes risk alleles, e.g., *TCF7L2* variants, and body mass index (BMI) differ in samples either enriched or depleted for type 2 diabetes (33).

**Case-only bias**

Inclusion is restricted to those with a disease (cases), so the analysis conditions on case status. This causes independent factors that determine case status to become statistically associated within the sample, even if these associations do not exist in the underlying population (34,35).

This review focuses on selection bias in GWAS. Selection bias in GWAS can propagate into downstream Mendelian randomization (MR), PGS and genetic correlation analyses, making reliable GWAS estimates essential.

A range of statistical approaches have been developed to address selection bias in human genetic studies, drawing on developments in epidemiology, econometrics and statistical genetics (36,37). We include methods developed for, or applied to, the estimation of SNP–trait associations in GWAS and organise them according to a practical decision framework: first, whether the analysis is diagnostic or corrective; second, within correction, what data are available and which form of selection is operating. This framework follows the decisions researchers face when addressing selection bias in GWAS data (Figure 1).

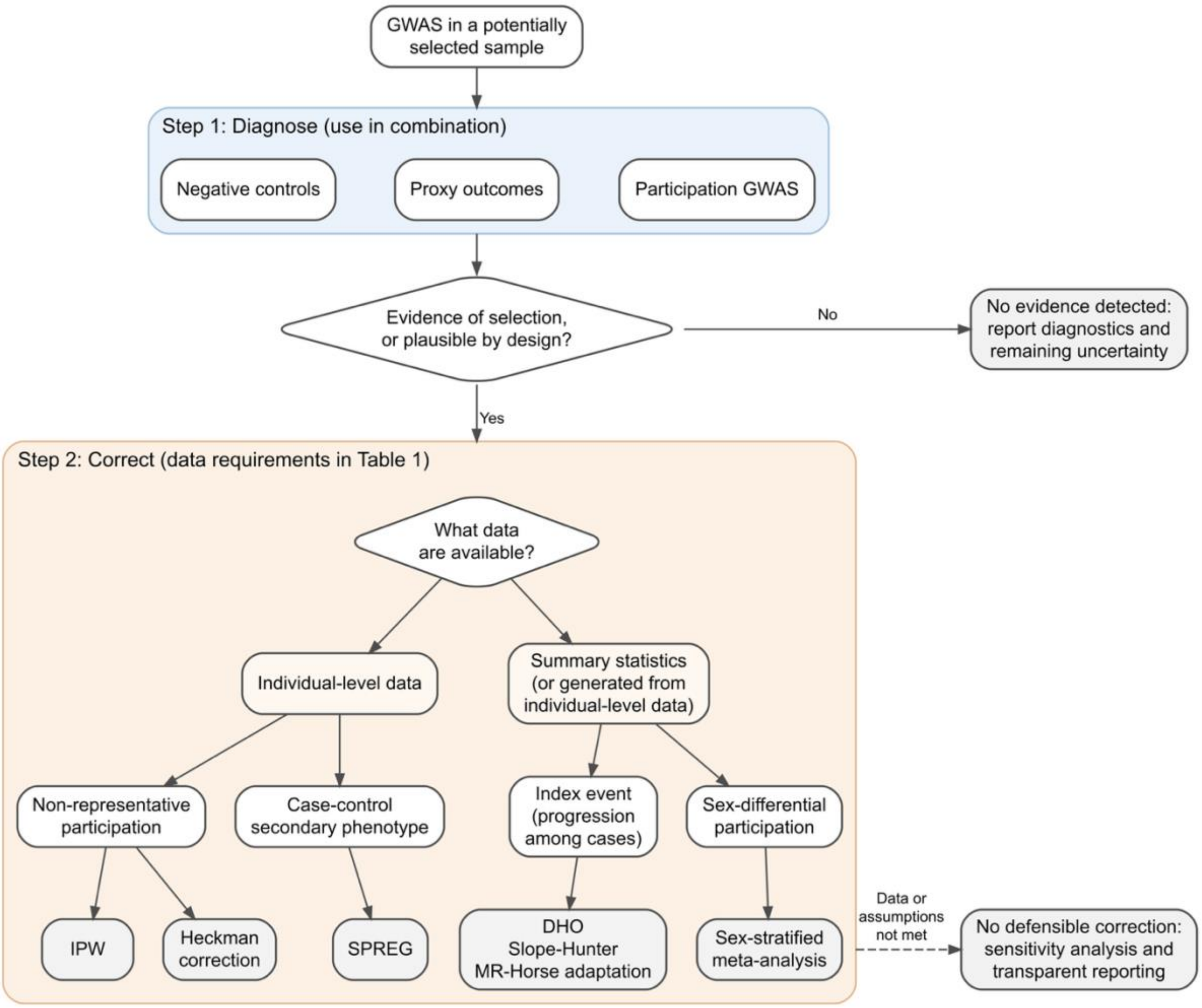

*Figure 1: A decision framework for addressing selection bias in GWAS. Step 1, diagnose: use complementary diagnostics with any data. Step 2, correct: choose a method according to data availability and the form of selection. Individual-level data can also be used to generate summary statistics.*

*DHO; Dudbridge Hedges-Olkin, GWAS; genome-wide association study, IPW; inverse probability weight, SPREG; Regression Analysis of Secondary Phenotype Data in Case-Control Association Studies*

For each method we describe the principle, its assumptions, the conditions under which it is and is not appropriate, and its data requirements. To help researchers implement these methods, we have developed a companion GitHub repository containing R implementations of each method applied to simulated data with known true effects for each selection structure, Figure 2.

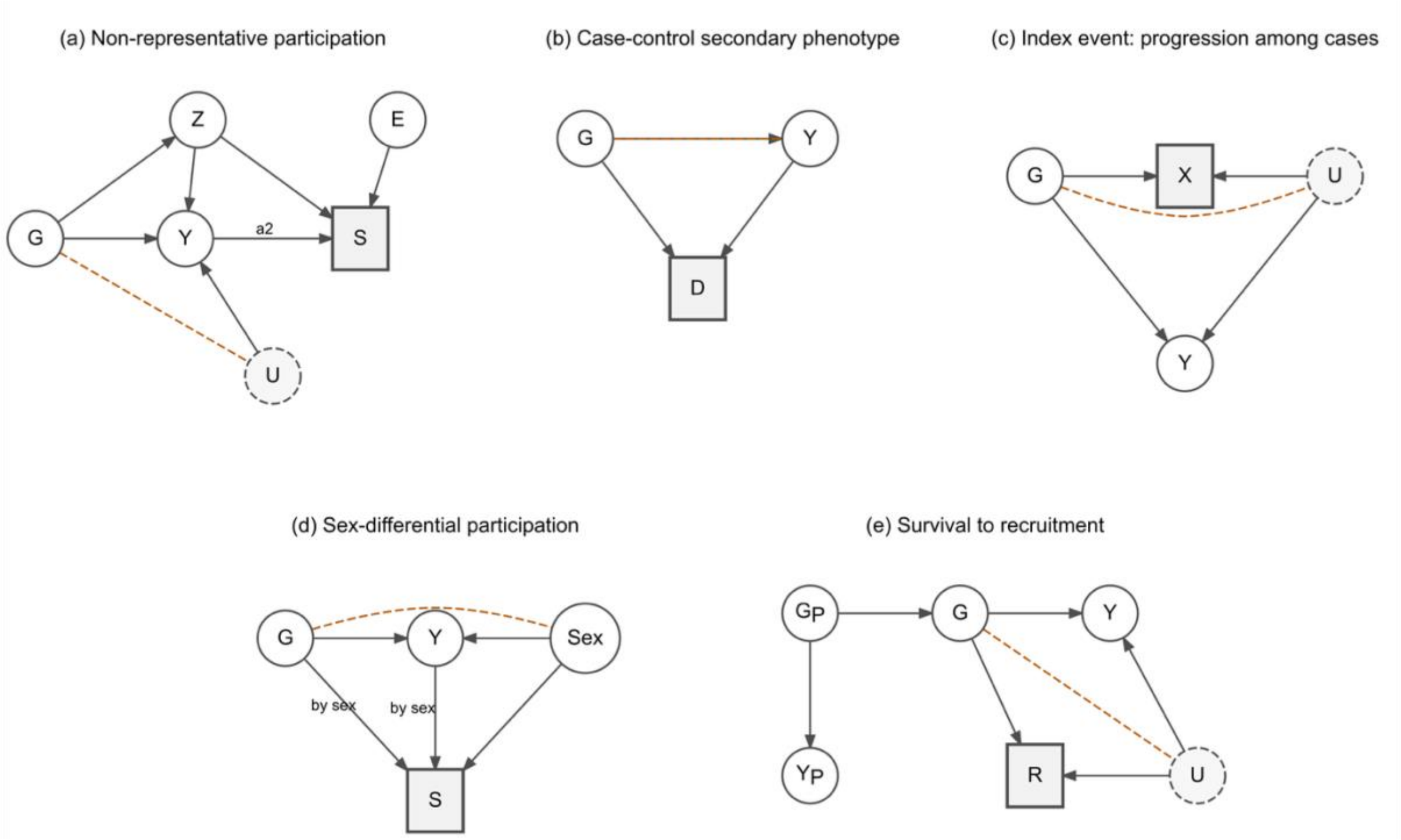

*Figure 2: Directed acyclic graphs (DAGs) of the selection structures addressed in this review. G, genotype; Y, trait; S, selection; U, unmeasured factors (dashed circles); Z, measured heritable predictor of selection; E, variable predicting selection but not Y; D, disease; X, disease incidence; R, alive at recruitment; $G_P$ and $Y_P$, parental genotype and outcome. Boxes denote conditioning; orange dashed lines denote associations induced by conditioning. (a) Non-representative participation (a1: selection on Z and E; a2: also on Y). (b) Case-control sampling with a secondary phenotype. (c) Index event: progression studied among cases. (d) Sex-differential participation: genetic and outcome effects on participation differ by sex. (e) Survival to recruitment, with parental outcomes as a proxy. Negative controls (phenotypes or variants with no causal link to G) would show induced associations in any panel. Each panel corresponds to a simulation module in the companion repository.*

Technical details for the methods are available in Supplementary Note 1. Code to generate the simulated data, method implementations and documentation are available at https://github.com/Ciarrah/gwas-selection-bias. Our interactive Shiny application provides guidance on method selection at https://oue54n-ciarrah-barry.shinyapps.io/app-1/.

This review does not cover SNP ascertainment bias, discussed elsewhere (10), nor collider bias induced by covariate adjustment, previously discussed (38–40). This review serves as an educational resource for researchers and practitioners seeking robust strategies to recognise, quantify, and mitigate selection bias in GWAS.

# Methods

## Diagnostic

Selection bias can be detected using the following approaches before proceeding to correction. These diagnostic methods apply regardless of downstream data availability and should be considered in any GWAS analysis using non-random samples. No single diagnostic is conclusive: a positive result may have causes other than selection, and a null result does not rule out selection bias. Diagnostics are therefore most informative in combination.

### *Negative controls*

Negative control outcomes are phenotypes for which no plausible biological association with the SNPs under investigation exists. Systematic inflation of SNP–negative control associations in the selected sample - relative to the full population or to external estimates - signals selection rather than biology (41,42). For example, in a GWAS of cardiovascular traits, significant associations (i.e., $p < 5 \times 10^{-8}$) between cardiovascular SNPs and an unrelated phenotype such as hair colour would not be expected; significant associations suggest selection is inducing spurious correlations. The genomic inflation factor (λ) computed for negative control outcomes provides a quantitative summary; λ substantially greater than 1.0 in the selected sample, absent in an external reference, indicates selection-induced inflation (43,44).

The negative control outcome is assumed independent of all SNPs in the study, not associated with loci in linkage disequilibrium (LD) with them, and free from pleiotropic pathways to the outcome of interest. The choice of negative control requires subject-matter justification in each setting and cannot be fully verified.

Negative controls are not appropriate when pleiotropy is widespread across the genome, as it becomes difficult to identify a truly unrelated phenotype. Implementation is provided in our companion GitHub repository.

### *Proxy outcomes*

When survival into the study sample is required to observe an outcome, estimates in the selected sample are restricted to survivors, potentially inducing survival bias. Familial genotypes can provide information about otherwise unobservable heritable traits through related individuals (45). For example, using parental disease status as a proxy outcome for participants' own disease removes the survival requirement, as the outcome is observed through a relative who need not themselves be a study participant. This reduces the selective exclusion of participants who died before recruitment and thereby mitigates survival-related selection bias (46). The participant's own outcome and the proxy outcome can each be analysed by GWAS and the results may be compared. After rescaling for relatedness (a parental proxy captures roughly half the participant's genetic effect), systematic deviation suggests survival-related or participation-related selection on the direct outcome.

Valid proxy-outcome inference assumes no measurement error or systematic missingness in familial outcome reporting. No alternative pathways should link the SNP and the proxy outcome through population structure, cohort effects, or age-related effect modification - for instance, if allele frequencies vary across age groups and selection into the sample is age-dependent (7).

Proxy outcomes are less informative when the trait has low familial correlation (limiting proxy validity and statistical power), or when cohort and age effects create systematic differences between the participant's and the proxy relative's exposure history. Proxy-outcome analyses should be reported alongside direct-outcome analyses as a sensitivity check. Implementation code is provided in our companion GitHub repository.

### *Participation GWAS*

A GWAS conducted with the participation indicator (selected vs. not selected) as the outcome directly identifies SNPs that predict selection into the sample. Significant SNP-participation associations indicate that genotype predicts study inclusion, which can distort SNP-trait associations in the selected sample (8,47). The proportion of SNPs reaching genome-wide significance should be compared against the expected false-positive rate; substantial excess signals genetic selection. Where neither exists, sex-differential participation can be detected among participants alone by a GWAS of sex, since autosomal variants cannot cause sex (6).

Participation indicators are typically unavailable in biobank settings where only participants are recruited but may be constructed through linkage to population registers. This diagnostic is not appropriate when the comparator is poorly characterised or when the proxy participation variable itself introduces measurement error that could inflate apparent selection signals. Standard GWAS software (PLINK, REGENIE) enables implementation and guidance for this is provided in our companion GitHub repository (48,49).

## Correction: individual-level data

These methods need individual-level genotype and phenotype data. IPW and the Heckman correction address non-representative participation. SPREG addresses case-control ascertainment of secondary phenotypes.

### *Inverse Probability Weighting (IPW)*

IPW adjusts a selected sample to reflect the target population. Each individual's ($i$) probability of inclusion, $P(S_i = 1| \boldsymbol{Z}_i)$, $S_i$ indicating sample inclusion and $\boldsymbol{Z}_i$ denoting covariates used to predict selection, is calculated. The inverse of the probability of being sampled is the weight, $w_i$, for the individual (50,51):

*[1]* $$w_i = \frac{1}{P(S_i=1| \boldsymbol{Z}_i)}$$

Underrepresented individuals are upweighted, and overrepresented individuals are downweighted, so that the distribution of observed characteristics in the weighted sample more closely reflects that in the population. SNP-trait associations are then estimated by weighted regression and are unbiased if the weighting model is correct. Weighted GWAS need robust (sandwich) standard errors. Extreme weights reduce the effective sample size, so stabilised or truncated weights are commonly used (37).

Several approaches have been used to estimate IPW weights (52–54). Schoeler et al. used auxiliary variables to generate sampling weights for UKB GWAS (52). The distribution of reweighted variables closely matched UK census data, indicating improved population representation. Differences between weighted and unweighted

GWAS estimates, including novel significant SNP-trait associations, suggested that selection bias may have affected the unweighted analyses.

IPW has been extended to correct for ascertainment bias in case-control studies when analysing multiple continuous secondary phenotypes, using scaled marginal models (SMM), Supplementary Note 1 (55). A common SNP effect across standardised phenotypes is estimated by weighted generalised estimating equations, which gains precision when the phenotypes share genetic effects.

Fundamentally, IPW assumes all variables determining study participation are correctly identified, adequately measured and modelled such that the weighting model and true underlying participation process correspond - this cannot be empirically verified (37). All population individuals must have non-zero probability of inclusion.

IPW is not appropriate when key selection covariates are unmeasured or when positivity is violated, i.e., when certain subgroups had effectively zero probability of entering the sample, as this causes weights to become unstable or infinite. Where selection is driven primarily by unmeasured genetic or phenotypic factors, IPW will numerically alter estimates which may give a false sense of bias removal. In such settings, the Heckman correction, which can accommodate selection on unmeasured factors under its modelling assumptions, may be considered alongside or instead of IPW (see below) (55,56). IPW implementation is provided in our companion GitHub repository.

### *Heckman*

The Heckman correction is an econometric approach to address bias from non-random sampling, jointly modelling the selection process and outcome to allow estimation of bias-adjusted SNP-trait associations (6,36,57). It uses properties of the normal distribution to adjust for the selection mechanism that generated the sample (Box 2).

*Box 2: Selection and outcome equations for the Heckman correction.*

The Heckman correction combines two equations: a selection equation modelling participation and an outcome equation modelling the relationship between the outcome and explanatory variables while accounting for selection:
Selection equation, fitted on sample participants and non-participants:

*[2]* $$P(S_i = 1 | X_i, Z_i) = \Phi(\delta X_i + \varphi Z_i)$$

Outcome equation, fitted only on participants:

*[3]* $$Y_i = \beta X_i + \rho\sigma_\epsilon \tau_i + \epsilon_i$$

with selection mechanism $(S)$, predictor variables including the genotype$(X)$, exclusion variables affecting $S$ not $Y(Z)$, vectors of coefficients for $X$ $(\delta)$ and $Z$ $(\varphi)$ and cumulative distribution function $(\Phi)$ in the selection equation; GWAS outcome of interest $(Y)$, the inverse Mills ratio $(\tau)$, selection bias corrected effect estimate of $X$ on $Y$ $(\beta)$, error term for the outcome model $(\epsilon)$, and selection parameter $(\rho\sigma_\epsilon)$ in the outcome equation. The method therefore has two linked steps: modelling participation, then modelling the outcome while accounting for selection.

Mignogna et al. applied a two-step Heckman correction to a GWAS of fluid intelligence measured at UK Biobank follow-up, using factor scores for questionnaire item nonresponse to model missingness (57). After correction, 3 of 11 genome-wide significant loci remained, lead-SNP effects were on average 15% smaller, and genetic correlations with other traits changed, for example with coronary artery disease.

The Heckman correction assumes the selection model is correctly specified. Identification requires an exclusion restriction: at least one variable that predicts selection but has no direct effect on the trait, otherwise the inverse Mills ratio becomes highly collinear with the outcome covariates, giving unstable estimates. The errors of the selection and outcome equations are assumed to follow a bivariate normal distribution with constant variance and a constant correlation, meaning selection on unobserved factors acts only through this correlation. The outcome model is assumed linear, with additive SNP effects. Selection predictors must be observed for both selected and non-selected individuals. When the selection equation is fitted once and its inverse Mills ratio is applied to every SNP, individual SNPs are additionally assumed to have negligible effects on selection.

The correction is most feasible for missingness within a cohort, such as attrition or nonresponse at follow-up, because covariates and genotypes are then available for both responders and non-responders. It is far harder for baseline participation, where non-participants are unobserved. Results are contingent on how the selection model is specified, so alternative specifications should be compared in sensitivity analyses. Software is available in the R package “sampleSelection”, and implementation code is provided in our companion GitHub repository. IPW and the Heckman correction thus address different drivers of participation, measured factors and the outcome or unmeasured factors respectively; their data requirements are compared in the Discussion.

### *Regression Analysis of Secondary Phenotype Data in Case-Control Association Studies (SPREG)*

Lin et al. propose retrospective likelihood modelling to address collider bias induced in secondary phenotype analysis within case-control studies, “Regression Analysis of Secondary Phenotype Data in Case-Control Association Studies” (SPREG)(58). SPREG models the joint distribution of the secondary phenotype and genotype, conditional on disease status; incorporating the sampling design into the likelihood removes collider bias. The probability of the genotype and the secondary trait are modelled conditional on disease status, accounting for the skewed phenotype distribution inherent to the outcome-based sampling approach.

He et al. extend SPREG using a Gaussian copula, allowing bias-corrected estimation of associations with secondary phenotypes under more flexible distributional assumptions (Supplementary Note 1) (59).

SPREG and its extension assume an additive genetic model and correct specification of the retrospective likelihood model. SPREG assumes the continuous secondary phenotype follows a normal linear regression model conditional on genotype and covariates. The Gaussian copula extension assumes each variable's latent (or transformed) representation follows a multivariate normal distribution.

SPREG is not appropriate for population-based or volunteer cohorts in which recruitment was not conditioned on disease, because the retrospective likelihood would be misspecified. It is also not suited to case-only designs. Where the secondary phenotype is markedly non-normal, the Gaussian copula extension should be used. Online software is available for implementation (60). For the same design, IPW and its scaled marginal model extension (see above) offer a more robust alternative, trading efficiency for fewer modelling assumptions (see Disucssion).

## Correction: summary-level data

When only GWAS summary statistics are available, corrections cannot model individual participation directly and instead rely on structural assumptions about how selection acts across many SNPs. Sex-stratified meta-analysis addresses sex-differential participation in population cohorts. The Dudbridge-Hedges-Olkin (DHO) correction, Slope-Hunter and the MR-Horse adaptation address index event bias in GWAS of disease progression, severity or survival among cases. Summary-level options for non-representative participation more generally are lacking.

### *Sex-stratified GWAS meta-analysis*

Study participation in large biobanks may systematically differ by sex, biasing or producing spurious association estimates. For example, Pirastu et al. found autosomal SNPs associated with sex in UKB and 23andMe (6). Such associations should not exist in the population, and they arise when the genetic determinants of participation differ between men and women. When participation depends on both sex and the trait under study, sex acts as a collider. Lee and Han show that, under low participation rates, sex-stratified GWAS estimates followed by a fixed-effects meta-analysis can correct for sex-differential participation bias (61). Within each stratum sex is constant, and the remaining bias in stratum-specific estimates is negligible when participation rates are low. The stratum estimates are combined by standard inverse-variance-weighted fixed-effects meta-analysis, with no loss of power relative to a combined GWAS in an unselected sample.

This approach assumes a logistic model of study participation and a low participation rate.

Sex-differential participation may be suspected when a GWAS of sex shows excessive signal or traits show a genetic correlation with sex. This correction is not appropriate as a general correction. It removes only bias acting through sex and leaves other selection mechanisms intact, so where those are also suspected it should be combined with other methods. The bias increases with participation rate, therefore this correction may not hold in cohorts with high participation. In principle, the same stratification argument applies to any binary or discrete variable affecting participation, although applications to date have focused on sex (61). Where SNP effects genuinely differ between sexes, sex-specific estimates should be reported alongside the combined estimate.

Running the stratified GWAS requires individual-level data, but published sex-stratified summary statistics are sufficient for the correction. Meta-analysis can be performed using standard software such as METAL. Implementation code is provided by the authors and in our companion GitHub repository (62,63).

### *Dudbridge-Hedges-Olkin (DHO)*

Studies of disease progression include only people with the disease, which can introduce index-event bias. Any SNP that raises the risk of incidence then becomes associated, among cases, with the other causes of incidence. If those causes also affect progression, the SNP acquires a spurious association with progression. This is index-event bias. For example, in a GWAS of recurrent events among people with a first myocardial infarction, variants that increase first-event risk may appear protective against recurrence. Cases carrying those variants needed fewer other risk factors to have a first event and so have lower risk of recurrence.

The Dudbridge Hedges-Olkin (DHO) method separates each SNP's observed effect on progression into its true effect and a bias proportional to its effect on incidence (35). Under the following linear additive models of disease incidence $X$ (i.e., the determinant of selection $S$), and progression $Y$:

*[4]* $$X = \beta_{GX}G + \beta_{UX}U + \varepsilon_X$$

*[5]* $$Y = \beta_{GY}G + \beta_{UY}U + \beta_{XY}X + \varepsilon_Y$$

with normally distributed random errors ($\varepsilon$), unmeasured common causes of incidence and progression ($U$), true direct SNP effect on progression ($\beta_{GY}$) and SNP effect on incidence ($\beta_{GX}$). The observed SNP effect on progression among cases ($\beta *_{GY}$) is biased by an amount proportional to the SNP effect on incidence:

*[6]* $$\hat{\beta}_{GY} = \hat{\beta} *_{GY} - \hat{b}\hat{\beta}_{GX}$$

where the bias slope is ($b$):

*[7]* $$b = -\frac{var(U)\beta_{UX}\beta_{UY}}{var(U)\beta_{UX}{}^2 + var(\varepsilon_X)}$$

As $U$ is unobserved, the quantities in [7] cannot be measured directly. Instead, $b$ is estimated from the data by weighted regression of the observed progression effects ($\hat{\beta} *_{GY}$) on the incidence effects ($\hat{\beta}_{GX}$) across independent SNPs. A Hedges–Olkin correction removes the regression dilution caused by sampling error in $\hat{\beta}_{GX}$. The estimated bias is then subtracted from each SNP's observed effect to recover the corrected estimate ($\hat{\beta}_{GY}$).

The DHO assumes linear and additive SNP effects, a constant bias slope across SNPs, and that direct SNP effects on progression are independent of SNP effects on incidence, attributing any observed correlation to bias. It may be applied with independent or overlapping samples. It is not appropriate when incidence and progression share biology, because the estimated slope is then biased and the correction incomplete; Slope-Hunter and the MR-Horse adaptation relax this assumption. An R package and documentation are available from the authors. We also provide an example implementation in our companion GitHub repository (64,65). A one-sample MR extension exists, but it targets covariate-induced collider bias in causal estimates and is outside the current scope (66).

### *Slope-Hunter*

Slope-Hunter addresses the same index-event setting as DHO but relaxes its independence assumption (34). SNPs are grouped by model-based clustering according to their effects on disease incidence and prognosis. SNPs that appear to affect incidence only are identified, and their pattern of association is used to estimate $b$ as any association between these SNPs and prognosis must arise from bias. This estimate is used to adjust the prognosis effects of all SNPs, as in [6]. Because $b$ is estimated from incidence-only SNPs, the remaining SNPs may have correlated effects on incidence and prognosis.

SNP effects are assumed approximately linear. Identification requires that incidence-only variants contribute at least as much to incidence heritability as pleiotropic variants, which cannot be tested directly.

Slope-Hunter may be the more appropriate of the two when incidence and prognosis plausibly share biology. It is not appropriate when the incidence-only cluster is small or poorly separated, because $b$ is then imprecisely estimated. It has shown inflated type I error in several simulation settings so results should be compared with DHO and the MR-Horse adaptation (67). An R package has been created by the authors and we provide an example implementation in our companion GitHub repository (68).

### *MR-Horse adaptation*

MR-Horse is a Bayesian MR method that accounts for widespread horizontal pleiotropy through a sparsity-inducing horseshoe prior (69). It has been adapted to correct GWAS of disease progression for index-event bias, modelling the association between SNP-incidence and SNP-progression effects among cases (70). In this adaptation, the bias slope $b$ ([7]) takes the place of the causal effect in MR, and each SNP's direct effect on prognosis ($\beta_{GY}$) takes the place of pleiotropy. The horseshoe prior shrinks most direct effects towards zero while allowing a few to be large. The posterior estimate of each direct effect is the bias-corrected prognosis effect.

The model assumes approximately linear and additive SNP effects, no effects of SNPs on U, and no effect modification. Unlike the DHO estimator, it does not require direct effects on prognosis to be independent of incidence effects, and unlike Slope-Hunter, it does not require a distinct incidence-only cluster. Instead, it assumes direct effects on prognosis are sparse: most SNPs have negligible direct effects, and a minority may have substantial ones.  LD is typically handled by pruning SNPs prior to analysis.

The MR-Horse adaptation is applicable in the index-event setting when neither independence (DHO) nor a well-separated incidence-only cluster (Slope-Hunter) is plausible, but sparsity of direct effects is. It is not appropriate when many SNPs are expected to have moderate direct effects on prognosis, which violates the sparsity assumption. Its computational cost also limits genome-wide application without pruning. It requires prior distributions, including a hierarchical horseshoe prior on direct SNP-prognosis effects, and a weakly informative prior on the collider-bias slope. Default priors are provided, but sensitivity analyses using alternative priors are recommended. Posterior inference is performed via Markov chain Monte Carlo in JAGS (71). We have provided example implementation in our companion GitHub repository. The three index event corrections thus differ mainly in their assumptions about pleiotropy; the Discussion summarise when each is preferable.

## Companion resources

## Repository

The companion GitHub repository (https://github.com/Ciarrah/gwas-selection-bias/) provides R implementations of nine of the methods described in this review, with input preparation for SPREG, organised in the same order as the text: diagnostics, individual-level corrections, summary-level corrections. Each method has a script, and each selection structure has a vignette that applies the relevant methods to simulated data, compares uncorrected and corrected estimates against the known true effects, and documents the required inputs and software. Installation instructions, the mapping of methods to scripts, and package versions are given in the repository README.

### *Worked example*

Consider a GWAS of a trait measured in a follow-up questionnaire completed by 30% of a genotyped cohort, where response depends on education (heritable, and affecting the trait), mode of invitation (affecting response only) and on the trait itself (Figure 2a, scenario a2). Non-responders are genotyped, meaning a participation GWAS is possible and shows strong inflation ($\lambda$ = 2.43). A negative control outcome shows little genome-wide inflation ($\lambda$ = 1.31) but strong inflation among education-associated variants ($\lambda$ = 19.0), so genome-wide summaries alone would understate the problem. Both IPW and the Heckman correction are feasible (see Discussion). IPW, weighting on education and mode of invitation , reduces the effective sample size from 15,090 to 819 and leaves the bias largely unchanged, because response also depends on the trait. Two checks guide the choice: the inverse Mills ratio coefficient in the Heckman model (0.63) indicates selection beyond education and mode of invitation, and mode of invitation is a credible exclusion variable only if it does not affect the trait directly, which cannot be empirically tested. The Heckman correction, using mode of invitation as the exclusion variable and including each variant in the selection equation, recovers the direct effects of trait-associated variants (bias -0.1%, against -14% for a naïve analysis adjusted for education, and -16% when variants are omitted from the selection equation), Figure 3.

*Figure 3: Worked example: non-response that depends on the outcome (Figure 2a, scenario a2). Estimated against true direct single nucleotide polymorphism (SNP) effects on the trait, with 95% confidence intervals (95% CI), for a naive analysis adjusted for education (left) and for the Heckman correction without (centre) and with (right) each SNP*

*in the selection equation. Points on the dashed line are unbiased. Slopes are the regression of estimated on true effects for trait-acting SNPs (1 = unbiased). Points at a true effect of zero are offset horizontally for visibility.*

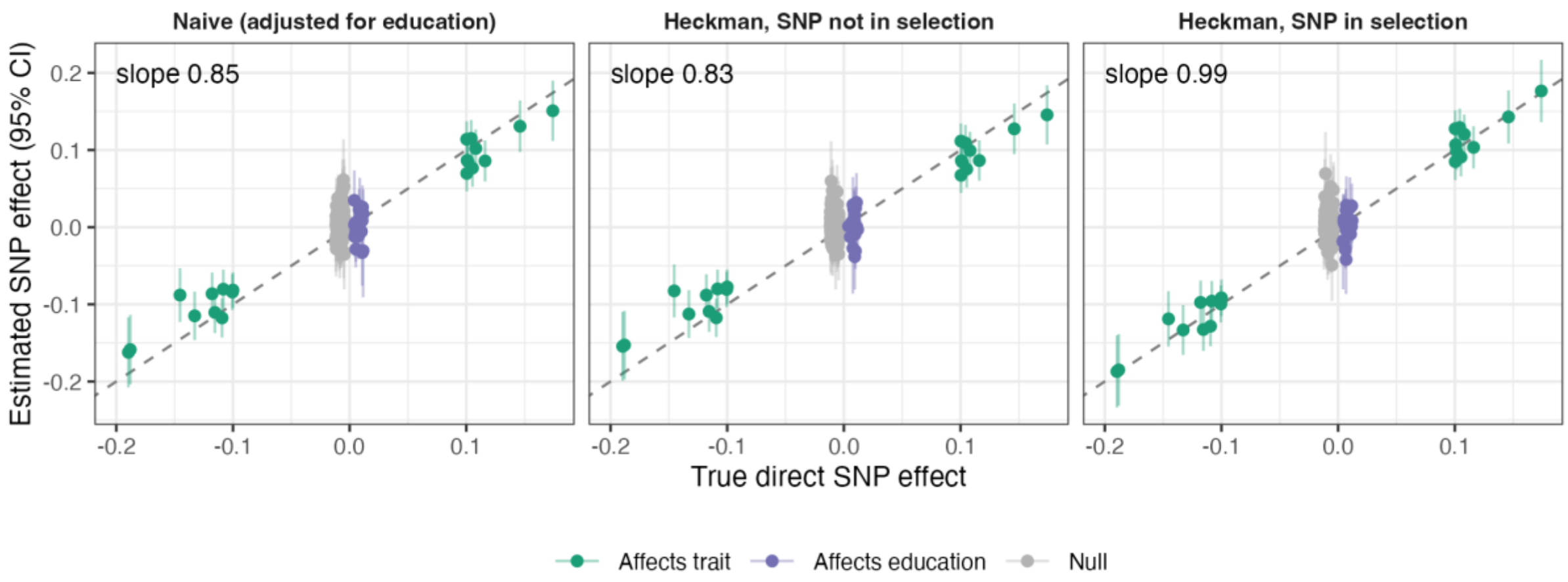

The other four selection structures in Figure 2 have equivalent simulation modules and vignettes in the repository. Further simulation details are available in Supplementary Note 2.

## Shiny app

The interactive application (https://oue54n-ciarrah-barry.shinyapps.io/app-1/) follows a decision framework. Users select their data availability and the form of selection they suspect, and the application returns applicable methods, their key assumptions, and links to the corresponding repository scripts, original methodological papers and implementation software.

# Discussion

No single method addresses all forms of selection. Method choice depends on the selection structure (Figure 2) and available data (Figure 1); Table 1 summarises the options.

*Table 1: Methods for detecting and correcting selection bias.*

| Setting | Method | Data required | Key assumption | Preferable when: | Limitations |
|---|---|---|---|---|---|
| **Detecting selection (any structure)** | Negative controls | Individual-level or summary-level data on a negative control outcome or negative control variants | The control shares causes with selection but has no causal relation with the genotype (control outcome) or the trait (control variant) | A credible control exists; works with summary statistics | A null result does not exclude bias; power depends on how strongly the control shares causes with selection; with few control variants, test the expected direction of bias rather than $\lambda$ |
| | Participation GWAS | Individual-level data with a participation indicator, population reference allele frequencies, or sex (GWAS of sex among participants) | The indicator or reference is accurate and representative; for the GWAS of sex, autosomal variants are independent of sex in the population | Any route is available; the GWAS of sex when there is no indicator or reference | Identifies genetic predictors of participation, not their effect on a given trait; the GWAS of sex detects only sex-differential participation and |
| | | | | | must exclude technical artefacts |
| | | Proxy outcomes | Individual-level genotypes with familial (e.g. parental) outcome data | Proxy outcomes are unaffected by the participant's selection; shared environment is negligible | Survival or mortality-related selection |
| **Non-representative participation** | IPW | Individual-level data plus non-participant data, a representative reference sample, or external population distributions (e.g. census) | Selection is fully explained by measured covariates (missing at random), with a correctly specified selection model | The drivers of participation are measured in both the sample and the reference | Cannot correct selection on the outcome or unmeasured factors; extreme weights sharply reduce the effective sample size |
| | Heckman correction | Individual-level data on participants and non-participants, including genotypes if variants affect participation | Bivariate normal errors in the selection and outcome equations; | Selection plausibly depends on the outcome or unmeasured factors, and non-participant | Without a credible exclusion variable, identification rests on distributional assumptions |
| | | | a valid exclusion variable | data exist (e.g. nonresponse within a genotyped cohort) | |
| **Case-control secondary phenotypes** | IPW / scaled marginal model | Individual-level data with disease status and population prevalence (or sampling fractions) | Known sampling fractions | Robustness is the priority; several secondary phenotypes share genetic effects (the scaled marginal model pools them) | Less efficient than likelihood-based methods |
| | SPREG | Individual-level data with disease status and population prevalence | A correctly specified joint model of disease and the secondary phenotype | Efficiency matters and the disease model is credible | Misspecification biases estimates; external software (64-bit Linux) |
| **Sex-differential participation** | Sex-stratified meta-analysis | Summary-level sex-stratified GWAS | Participation is rare enough that selection acts multiplicatively within each sex | A GWAS of sex suggests sex-differential participation | Does not remove bias when participation is common |
| **Index event (progression among cases)** | DHO | Summary-level incidence and progression GWAS | Effects on progression are uncorrelated with effects on incidence across variants | A first-line baseline: fast, uses all independent variants | Under-corrects, and can reverse the correction, under correlated pleiotropy |
| | Slope-Hunter | Summary-level incidence and progression GWAS | Incidence-only variants contribute at least as much to incidence heritability as pleiotropic variants | Correlated pleiotropy is plausible and many incidence-only variants exist | Can identify the wrong cluster when incidence-only variants are few or weak, inflating type I error |
| | MR-Horse adaptation | Summary-level incidence and progression GWAS | Direct effects on progression are sparse; pleiotropy may be correlated with incidence effects | Correlated pleiotropy and sparse direct effects are both plausible | Computationally heavier; its prior can over-correct when pleiotropy is uncorrelated |

Software and code for each method are listed in the companion repository (https://github.com/Ciarrah/gwas-selection-bias).

Methods suited to the same structure differ mainly in what they assume and in the data they need about non-participants. For non-representative participation, IPW corrects selection on measured factors, with weights estimated from non-participant data, a population-representative reference sample or external population distributions, but it cannot correct selection on factors not measured in both the sample and the reference. The Heckman correction can address selection on the outcome or unmeasured factors, but requires individual-level non-participant data, including genotypes when variants influence participation, and a credible exclusion variable, which is realistic for nonresponse within a genotyped cohort but rarely for selection into the cohort itself. For case-control secondary phenotypes, IPW and its scaled marginal model extension are robust, needing only sampling fractions, whereas SPREG is more efficient when its joint disease model is correct. For index events, the three corrections share a structure but differ in their assumptions about pleiotropy: DHO assumes effects on incidence and progression are uncorrelated and is a natural baseline; Slope-Hunter relaxes this by identifying incidence-only variants but performs poorly when these are few or weak; and the MR-Horse adaptation assumes sparse direct effects, at greater computational cost. Summary-level corrections rely on assumptions about the genetic architecture that cannot be fully tested, and some inflate false-positive rates (67). No method is robust to every violation, therefore applying methods with different assumptions and comparing results indicates whether a conclusion depends on a particular assumption.

Diagnostics are as important as corrections. Bias may be concentrated in variants linked to selection. Genome-wide summaries, such as the genomic inflation factor, may therefore appear unaffected even when individual estimates are substantially distorted. Diagnostics are most informative when examined for variants plausibly related to selection.

Selection also interacts with other sources of structure in genetic data. Assortative mating induces correlations between variants that GWAS assumes to be independent, particularly when relatives or couples are oversampled (72), and population stratification can induce spurious SNP-trait associations (73). Genetic principal components or relatedness-based random effects should be included routinely, alongside any selection correction.

We identify three strategies for future studies: prevention, sensitivity and transparency (Box 3).

*Box 3: Actionable guidance for addressing selection bias in genome-wide association studies (GWAS).*

**Prevention**

- Establish large and well-characterised population-representative samples as reference sets for more selective studies.
- Collect variables or proxies plausibly related to participation, follow-up and survival alongside those needed for the research question, to support later sensitivity analyses and bias correction.
- Where possible, link samples to population registries or census data to enable calibration of selection mechanisms.

**Sensitivity**

- Apply diagnostics, such as negative controls and participation GWAS, to detect selection and to check whether corrections have removed it.
- Compare results from relevant bias correction methods with different assumptions or data requirements to assess whether the findings converge.
- Compare evidence from cohorts, biobanks, case-control studies and other sources with different selection mechanisms to assess the consistency and range of estimates.

- For lead associations carried forward into downstream analyses, bounding approaches can quantify how strong selection would need to be to explain the result (74).

**Transparency**

- Systematically document the sample selection mechanism and inclusion/exclusion criteria.
- Report the assumptions underlying each bias correction method used.
- Describe the limitations and potential consequences of the bias correction methods used.
- More studies that model the key determinants of study participation and responsiveness are required.

Design-based prevention remains preferable to post-hoc correction but is resource-intensive, and underrepresentation can persist even in national samples (75). Where prevention is not possible, triangulation across data sources and methods with different selection mechanisms and assumptions provides a plausible range of estimates (76). Transparent reporting does not remove bias but enables it to be quantified and addressed across studies.

## Conclusion

Selection bias in GWAS is pervasive but often detectable, and in many settings correctable, provided the selection structure is understood and the required data are available. Population-wide data with known selection mechanisms are unlikely to become routine, so progress depends on bias-aware study designs, routine diagnostics, corrections with explicit assumptions and transparent reporting. Because GWAS results underpin downstream causal inference and risk prediction, addressing selection bias in the primary analysis also benefits downstream analyses.

## Funding

CSB is supported by the Pioneer Centre for Statistical and computational Methods for Advanced Research to Transform Biomedicine (SMARTbiomed), DNRF grant number P4. BJV is supported by the Independent Research Fund (2034-00241B) and the Lundbeck Foundation (R335-2019-2339). EA was funded by the Novo Nordisk Foundation in the development and use of the Multi-Generation Register (grant NNF20SA0066568). ZK is funded by the Swiss National Science Foundation (315230-219587).

## Data availability

No new data were generated or analysed from human participants. All simulated data can be regenerated with the code in the companion repository (https://github.com/Ciarrah/gwas-selection-bias; archived at Zenodo, , https://doi.org/10.5281/zenodo.23943773)).

## Ethics statement

Not applicable. This article reviews published methods and uses only simulated data; no human participants or animals were involved.

## Acknowledgements

In accordance with Briefings in Bioinformatics and Oxford Academic policies regarding artificial intelligence, we disclose that Claude (Anthropic) was used during the preparation of this submission to assist with reviewing and documenting the R code in our companion repository and Shiny application, as well as to check the manuscript for grammatical errors and inconsistencies. All scientific content, text, tables, and figures were generated by the authors, who reviewed and tested all code and take full responsibility for the final content.

## Declaration of interests

All authors declare no conflicts of interest.

## Availability of data and materials

No new datasets are generated within this review. The full search strategies, including keywords and database queries, are provided in the Supplementary Materials. Our interactive web application is publicly available at https://github.com/Ciarrah/gwas-selection-bias. Our interactive Shiny application is available for method selection guidance at https://oue54n-ciarrah-barry.shinyapps.io/app-1/.

## Authors contributions

**CSB**: Conceptualization, Data Curation, Investigation, Methodology, Software, Visualization, Writing – Original Draft. **TG**: Data Curation, Investigation, Validation, Visualization. **ZK:** Writing – Review & Editing. **BJV**: Funding Acquisition, Writing – Review & Editing. **EA**: Funding Acquisition, Supervision, Validation, Writing – Review & Editing.

# Supplementary Materials: Navigating Selection Bias in Genome-wide Association Studies (GWAS)

# Supplementary notes

## *Supplementary note 1: Technical details*

### Negative controls

For negative control outcomes in population cohorts (main text), inflation is summarised by the genomic inflation factor $\lambda = \frac{median(\chi_j^2)}{0.4549}$ where $\chi_j^2$ is the association test statistic for SNP $j$ with the negative control outcome, and 0.4549 is the median of a $\chi^2$ distribution with one degree of freedom (1). $\lambda$ should be compared with the same negative control analysed in an external reference sample as polygenicity and population stratification can also inflate $\lambda$.

### Proxy outcome

Under an additive model, the association between participant genotype and a parental outcome is approximately half the direct effect:

$$\beta_{G,parent} = \frac{\beta_{G,offspring}}{2}$$

Proxy estimates are therefore multiplied by two before comparison with direct estimates (2).

### Inverse Probability Weighting

Stabilised weights rescale each weight by the marginal probability of participation (3):

$$w'_i = \frac{P(S_i = 1)}{P(S_i = 1|\, Z_i)}$$

Scaled marginal model (SMM): for M standardised secondary phenotypes, the mean model is (4):

$$E(\boldsymbol{Y}_{\boldsymbol{i,j}}^{\boldsymbol{Scaled}} | \boldsymbol{Z}_i, G_i) = \boldsymbol{Z}_i^T \boldsymbol{\beta}_j^* + G_i \alpha_j^*$$

where $\boldsymbol{Y}_{\boldsymbol{j}}^{\boldsymbol{Scaled}}$ are correlated standardised phenotypes, $\boldsymbol{\beta}_j^*$ the phenotype-specific covariate effects, $\alpha_j^*$ the phenotype-specific biased SNP effects, and $G_i$ the variant genotype, with $M$ phenotypes and $N$ individuals. Rewritten in matrix form:

$E(\boldsymbol{Y}_{\boldsymbol{i}}^{\boldsymbol{Scaled}} | \boldsymbol{X}_{\boldsymbol{i}}) = \boldsymbol{X}_{\boldsymbol{i}} \boldsymbol{\gamma}$ where $\boldsymbol{Y}_{\boldsymbol{i}}^{\boldsymbol{Scaled}} = (\boldsymbol{Y}_{\boldsymbol{i,1}}^{\boldsymbol{Scaled}}, \ldots, \boldsymbol{Y}_{\boldsymbol{i,M}}^{\boldsymbol{Scaled}}), \boldsymbol{X}_{\boldsymbol{i}} = [\boldsymbol{Z}_{\boldsymbol{i}} \boldsymbol{G}_{\boldsymbol{i}}]$and $\boldsymbol{\gamma}$ the vector of scaled covariate and shared SNP effects across phenotypes. Applying IPW to the GEE we obtain:

$$\sum_{i=1}^{n} w_i \boldsymbol{X}_i^T \boldsymbol{R}^{-1} (\boldsymbol{Y}_{\boldsymbol{i}}^{\boldsymbol{Scaled}} - \boldsymbol{X}_i^T \gamma) = 0$$

where $\boldsymbol{R}$ is the correlation matrix of for the scaled phenotypes. Robust standard errors are computed for $\alpha$ at convergence.

## Heckman correction

The inverse Mills ratio (IMR) is computed from the fitted probit selection model:

$$\hat{\tau}_i = \frac{\phi(\hat{\delta} X_i + \hat{\varphi} Z_i)}{\Phi(\hat{\delta} X_i + \hat{\varphi} Z_i)}$$

With standard normal density ($\phi$) and distribution ($\Phi$) functions. $\hat{\tau}_i$ is an estimate, meaning second stage standard errors require correction (5).

### *Regression Analysis of Secondary Phenotype Data in Case-Control Association Studies (SPREG)*

Lin et al. observe covariates ($Z$), a secondary dichotomous phenotype ($Y$), and genotype ($G$) conditional on disease status ($D$)(6). Bayes rule is employed to factorise the observed joint distribution $f(G, Z, Y| D)$ into the underlying population model intended to be estimated, utilising the following relation:

$$f(G, Z, Y|D) \propto f(Y|G, Z) f(D|Y, G, Z)$$

The retrospective likelihood to minimise is:

$$\mathcal{L}(\beta,\sigma_Y^2) = \prod_{i=1}^{N} \frac{f(Y_i|\,\boldsymbol{Z}_i, G_i;\,\beta) f(D_i|\,\boldsymbol{Z}_i, G_i, Y_i\,;\,\delta)}{P(D_i)}$$

where $\beta$ parameterises the secondary phenotype model, $\delta$ the disease model and $P(D_i)$is evaluated from known population prevalence (6). The population distribution of $\boldsymbol{G}$ and $Z$ is profiled out.

He et al. form a multivariate distribution for the likely correlation between primary and secondary phenotypes of the case-control study, employing a copula to link the marginal distributions of multivariate random variables to their joint distribution(7,8). First, the marginal variables are transformed to have uniform distribution on [0, 1]:

$U_A = F(A)$ such that $U \sim Unif(0,1)$ and $A \in \{G, Z, Y\}$

The subsequent dependence structure is expressed as a multivariate distribution on these transformed variables, with correlation structure $R_D$ and $K$ the total number of variables being jointly modelled:

$Z_A = \Phi^{-1}(A)$ such that $(Z_X, Z_Y, Z_G|D) \sim \mathcal{N}_K(0, R_D)$

The copula (retrospective) likelihood is defined and optimised over $D$ to obtain bias corrected estimates:

$$\mathcal{L}(\beta,\sigma_Y^2) = \prod_{i,d}^{N,D_i} f(Z_{X_i}, Z_{Y_i}, Z_{G_i}\,|D, R_{D_i})$$

### *Sex-stratified GWAS meta-analysis*

Sex-specific estimates $\beta'_{GY}$(e.g., female) and $\beta'_{GY}$(e.g., male) are combined by inverse-variance-weighted (IVW) fixed-effects meta-analysis (9):

$$\hat{\beta}_{GY} = \frac{\hat{\beta}_{GY}^{F}/SE_F^2 + \hat{\beta}_{GY}^{M}/SE_M^2}{1/SE_F^2 + 1/SE_M^2}, \qquad SE^2(\hat{\beta}_{GY}) = \left(\frac{1}{SE_F^2} + \frac{1}{SE_M^2}\right)^{-1}$$

### *Dudbridge Hedges Olkin (DHO)*

The bias slope $b$ is estimated by weighted regression of $\hat{\beta} *_{GY}$on $\hat{\beta}_{GX}$across independent SNPs, with weights $w_i = \frac{1}{SE^2}\hat{\beta} *_{GY_j}$. The Hedges–Olkin correction for regression dilution is:

$$\hat{b} = \hat{b}_{obs} \times \frac{\sum_j w_j \hat{\beta}^2_{GX,j}}{\sum_j w_j \hat{\beta}^2_{GX,j} - \sum_j w_j \; SE^2(\hat{\beta}_{GX,j})}, \qquad w_j = \frac{1}{SE^2(\hat{\beta}^*_{GY,j})}$$

Alternatively, simulation extrapolation (SIMEX) may be used. The variance of each corrected estimate is approximately $\mathrm{var}(\hat{\beta}_{GY}) \approx SE^2(\hat{\beta}^*_{GY}) + \hat{b}^2 \, SE^2(\hat{\beta}_{GX}) + \hat{\beta}^2_{GX} \, \mathrm{var}(\hat{b})$ (10).

### *Slope-Hunter*

Among SNPs associated with disease incidence, a Gaussian mixture model is fitted in the $(\hat{\beta}_{GX}, \widehat{\beta*}_{GY})$ plane. The component whose points lie closets to the line through the origin is taken as the incidence-only cluster. The bias $b$ is estimated from this slope, corrected for regression dilution (11).

### *MR-Horse adaptation*

For variant i, the observed SNP-progression association among cases is modelled as the direct effect plus collider bias, and the observed SNP-incidence association as its true value plus sampling error (12).

$$\hat{\beta}^*_{GY,i} \sim \mathcal{N}(\beta_{GY,i} + b\,\beta_{GX,i},\ \sigma^2_{GY,i}), \qquad \hat{\beta}_{GX,i} \sim \mathcal{N}(\beta_{GX,i},\ \sigma^2_{GX,i})$$

The true effects on incidence and progression share a bivariate normal prior. A horseshoe prior is placed on $\beta_{GY,i}$, and $\rho_i$ allows the two effects to be correlated:

$$\begin{pmatrix} \beta_{GX,i} \\ \beta_{GY,i} \end{pmatrix} \sim \mathcal{N}\left( \begin{pmatrix} \mu \\ 0 \end{pmatrix}, \begin{pmatrix} s_x^2 & \rho_i s_x \phi_i \tau \\ \rho_i s_x \phi_i \tau & \phi_i^2 \tau^2 \end{pmatrix} \right)$$

where $\phi_i$ and $\tau$ are the local and global shrinkage parameters. Hyperpriors are:

$$\rho_i = 2r_i - 1,\ \ r_i \sim \mathrm{Beta}(3,1); \qquad \phi_i,\ \tau \sim \mathrm{Cauchy}^+(0,1)$$

$$b \sim \mathrm{Uniform}(-10,10); \qquad \mu \sim \mathcal{N}(0,1); \qquad s_x^2 \sim \mathcal{N}^+(0,1)$$

The $\mathrm{Beta}(3,1)$ prior places most weight on strong positive correlation between incidence and progression effects. Donovan et al. chose it for chronic kidney disease, so it should be adapted to the trait studied. Posterior estimates are obtained by Markov chain Monte Carlo in JAGS (13). The posterior estimate of $\beta_{GY,i}$ is the bias-corrected SNP-progression effect.

## Supplementary note 2: Simulations

Simulations were designed to enable illustration of each method; each scenario is a single replicate with a fixed seed. SNPs were independent, and performance was assessed separately for SNPs affecting the outcome, affecting selection, or neither. Full code, parameters and worked examples are in the companion repository.

| **Module (Figure 2)** | **Sample** | **Selection mechanism** | **Scenarios** | **Methods applied** |
|---|---|---|---|---|
| (a) Participation | 50,000; 200 SNPs | Probit participation on Z and E (a1), plus Y (a2); 30% participate | a1, a2 | Participation GWAS, negative control outcome, IPW, Heckman |
| (b) Case-control | 200,000 → 5,000 cases + 5,000 controls; 200 SNPs | Sampling on disease (K = 5%) | b | IPW, scaled marginal model, SPREG |
| (c) Index event | 50,000 (15,000 cases); 1,000 SNPs | Progression studied among cases | c1 independent, c2 correlated pleiotropy, c3 sparse | Negative control variants, DHO, Slope-Hunter, MR-Horse |
| (d) Sex-differential | about 15,000 participants; 200 SNPs | Logistic participation on G and Y, with effects differing by sex | d_low: 5% participation; d_high: 50% participation | Participation GWAS (via GWAS of sex), sex-stratified meta-analysis |
| (e) Survival | 200,000 (60,000 survivors); 200 SNPs | Survival on G and U | e | Proxy outcomes |

GWAS; genome-wide association study, MR; Mendelian randomization, SNP; single nucleotide polymorphism